\documentclass[%reprint,
superscriptaddress,
 amsmath,amssymb,
 aps,
twocolumn,
]{revtex4-2}
\usepackage{graphicx}% Include figure files
\usepackage{dcolumn}% Align table columns on decimal point
\usepackage{bm}% bold math
\usepackage{amssymb}
\usepackage{float}
\usepackage{hyperref}% add hypertext capabilities

\begin{document}

%%% Title %%%
\title{Space-time beams with tunable orbital group velocity toward plasma superradiance}

%%% Authors %%%%
\author{Gabrielle Vaz}
\thanks{These authors contributed equally to this work.}
\affiliation{GoLP/Instituto de Plasmas e Fus\~ao Nuclear,  Instituto Superior T\'ecnico, Universidade de Lisboa, Lisbon, Portugal}
\affiliation{Department of Physics, Instituto Superior T\'ecnico, Universidade de Lisboa, Lisbon, Portugal}
\affiliation{Instituto de Engenharia de Sistemas e Computadores – Microsistemas e Nanotecnologias (INESC MN) Lisbon, Portugal}

\author{Rafael Almeida}
\thanks{These authors contributed equally to this work.}
\affiliation{GoLP/Instituto de Plasmas e Fus\~ao Nuclear,  Instituto Superior T\'ecnico, Universidade de Lisboa, Lisbon, Portugal}
\affiliation{Department of Physics, Instituto Superior T\'ecnico, Universidade de Lisboa, Lisbon, Portugal}

\author{Pablo San Miguel Claveria}
\thanks{These authors contributed equally to this work.}
\affiliation{GoLP/Instituto de Plasmas e Fus\~ao Nuclear,  Instituto Superior T\'ecnico, Universidade de Lisboa, Lisbon, Portugal}
\affiliation{Department of Physics, Instituto Superior T\'ecnico, Universidade de Lisboa, Lisbon, Portugal}

\author{Robert Neumann}
\affiliation{GoLP/Instituto de Plasmas e Fus\~ao Nuclear,  Instituto Superior T\'ecnico, Universidade de Lisboa, Lisbon, Portugal}
\affiliation{Department of Physics, Instituto Superior T\'ecnico, Universidade de Lisboa, Lisbon, Portugal}

\author{Joaquim Pereira}
\affiliation{Department of Physics, Instituto Superior T\'ecnico, Universidade de Lisboa, Lisbon, Portugal}
\affiliation{Instituto de Engenharia de Sistemas e Computadores – Microsistemas e Nanotecnologias (INESC MN) Lisbon, Portugal}

\author{Carolina Miranda}
\affiliation{Department of Physics, Instituto Superior T\'ecnico, Universidade de Lisboa, Lisbon, Portugal}
\affiliation{Instituto de Engenharia de Sistemas e Computadores – Microsistemas e Nanotecnologias (INESC MN) Lisbon, Portugal}

\author{Vincent Ginis}
\affiliation{Data Analytics Lab, Vrije Universiteit Brussel, 1050 Brussel, Belgium}

\author{Jorge Vieira}
\email{jorge.vieira@tecnico.ulisboa.pt}
\affiliation{GoLP/Instituto de Plasmas e Fus\~ao Nuclear,  Instituto Superior T\'ecnico, Universidade de Lisboa, Lisbon, Portugal}
\affiliation{Department of Physics, Instituto Superior T\'ecnico, Universidade de Lisboa, Lisbon, Portugal}

\author{Marta Fajardo}
\email{marta.fajardo@tecnico.ulisboa.pt}
\affiliation{GoLP/Instituto de Plasmas e Fus\~ao Nuclear,  Instituto Superior T\'ecnico, Universidade de Lisboa, Lisbon, Portugal}
\affiliation{Department of Physics, Instituto Superior T\'ecnico, Universidade de Lisboa, Lisbon, Portugal}

\author{Marco Piccardo}
\email{marco.piccardo@tecnico.ulisboa.pt}
\affiliation{Department of Physics, Instituto Superior T\'ecnico, Universidade de Lisboa, Lisbon, Portugal}
\affiliation{Instituto de Engenharia de Sistemas e Computadores – Microsistemas e Nanotecnologias (INESC MN) Lisbon, Portugal}

\date{\today}

%%% Abstract %%%
\begin{abstract}
Light springs are space-time beams that have a helical wavepacket. Due to this special property, light springs result into a rotating pulse when intercepting a plane lying orthogonal to their propagation direction. Associated to this, we introduce here the orbital group velocity, an additional tunable property of light springs. The orbital group velocity quantifies the speed of the light spring intensity rotation, distinctly from the conventional longitudinal group velocity, which describes the motion of the wavepacket envelope along its propagation axis. We demonstrate experimentally by tunable Fourier synthesis that the orbital group velocity can assume sub- and superluminal values, thus becoming a new platform for synthetic motion studies and control of laser-matter interactions. Particularly, in the superluminal regime, when interacting with a thin overdense plasma, we reveal by particle-in-cell simulations that the light spring unlocks superradiant radiation, due to the coherent excitation of the electrons in the plasma acting as a quasiparticle. This superradiant source inherits the ultrafast temporal dynamics of the light springs while emitting in the terahertz region, thus creating a new source of terahertz radiation controlled by the properties of spatiotemporal coupling of the laser. Therefore, spatiotemporal tuning of light springs is at the frontier of controlling laser-matter interaction and generating new tunable sources of radiation.
\end{abstract}

%%% Main text %%%
\maketitle

%%% Introduction + Lighthouse %%%
Space-time couplings appear in optical systems when the spectral properties of a beam depend on position, resulting in complex spatiotemporal correlations in the light field~\cite{STNoise}. In ultrafast laser systems, these couplings were initially perceived as chromatic aberrations and often treated as a detrimental effect to be minimized~\cite{STNoise,STAberration}. However, in recent years, beams with tunable spatiotemporal properties have been synthesized~\cite{STBeams,STBeams1} and are unlocking the possibility to explore new regimes of light-matter interaction~\cite{STMultimodeShaping} with applications in fields such as nonlinear conversion~\cite{NLConversion1,NLConversion2} or quantum metrology~\cite{QuantumMetrology}. This class of optical beams includes propagation-invariant laser pulses~\cite{InvariantSTBeams, almeida_universal_2025} and spatiotemporal optical vortices~\cite{STOV}. A particularly fascinating feature of spatiotemporal beams in pulsed laser systems is the possibility to tune the longitudinal velocity of the intensity peak, referred to as the group velocity ($v_{g}$), to values above the vacuum light speed $c$, without violating relativistic causality \cite{AbouraddyGV,STwavepackets, sainte-marie_controlling_2017}, such as in the case of the flying-focus~\cite{Froula_PoP_2017,FlyingFocus, kabacinski_spatio-temporal_2023, liberman_use_2024, markland_rapidly_2026}. 

%%% Figure: Lighthouse
\begin{figure}[hb!]
\includegraphics[scale=0.3]{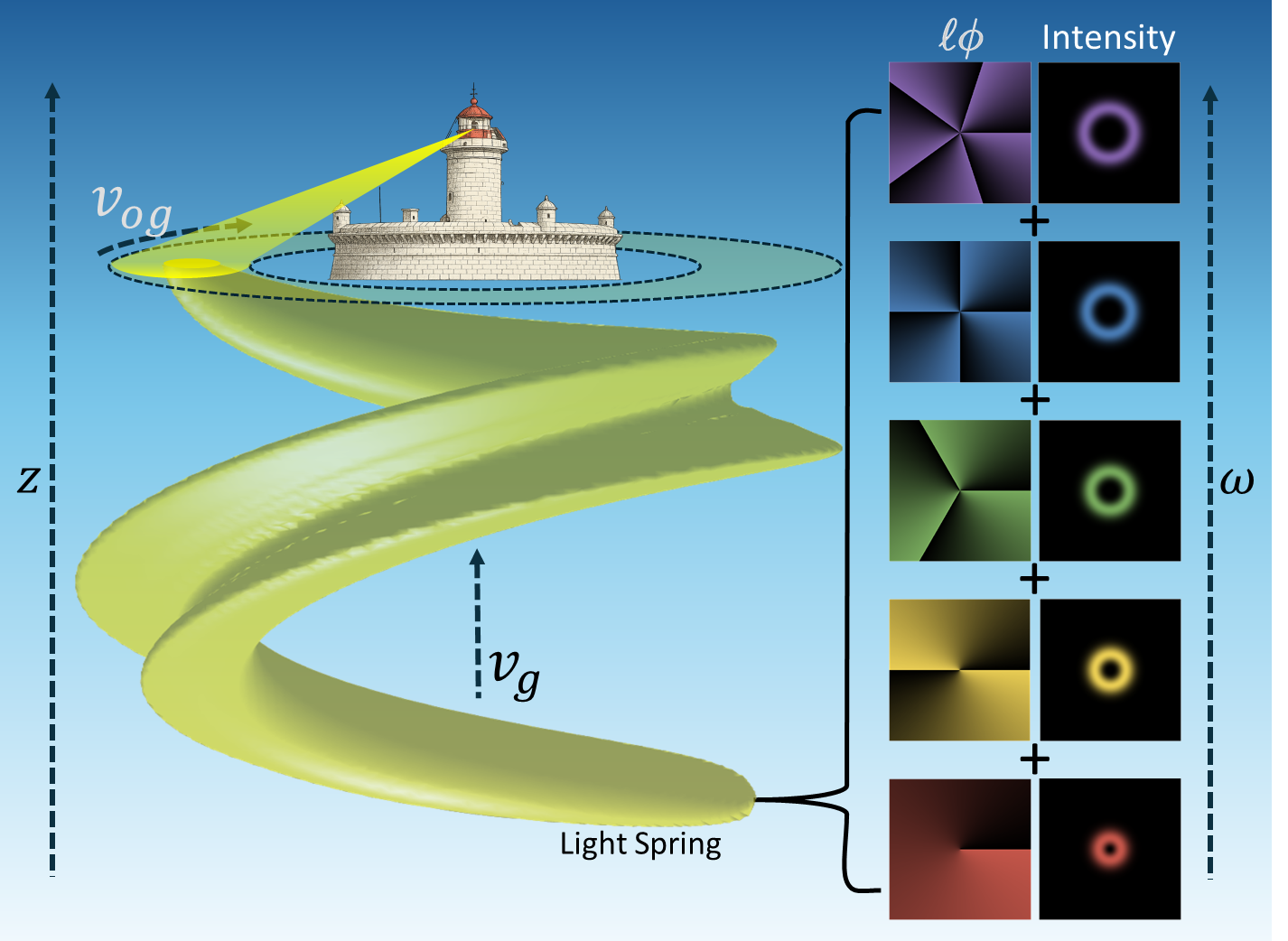}
\caption{\label{fig:Concept} \textbf{Lighthouse analogy for the orbital group velocity of a light spring.} The lighthouse is used here as a kinematic analogy for the apparent rotation of the intensity hotspot of a light spring---a space-time coupled beam created when different topological charges ($\ell$) are applied to different spectral components ($\omega$) in a broadband laser pulse. In both systems, the observed tangential speed is set by an angular sweep rate and an observation radius. The maximum intensity of the wavepacket rotates at a tangential speed $v_{og}$ on an intercepting plane, as the beam propagates with a longitudinal group velocity $v_{g}$. Similarly, the light emitted by a lighthouse produces the same pattern. $z$ is the propagation axis.}
\end{figure}

%%% Figure LS: simulation + experiments %%%
\begin{figure*}[ht] 
\includegraphics[scale=0.48]{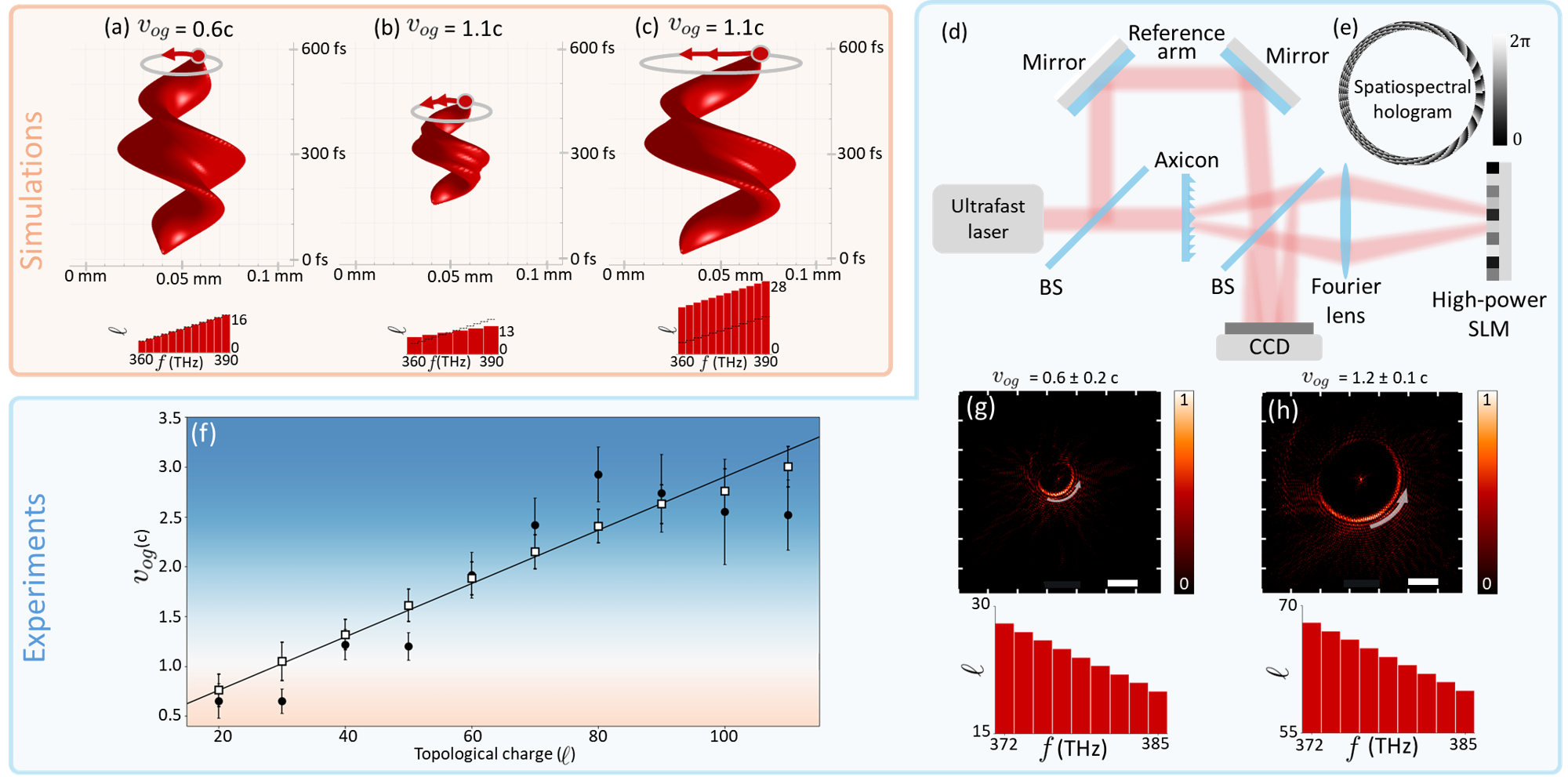}
\caption{\label{fig:OGV} \textbf{Tuning the orbital group velocity of light springs from sub- to superluminal.} The orbital group velocity, $v_{og}$, tunability of a light spring can be accessed through control over the spatiospectral correlation: by changing the slope of the linear relation---(a) to (b)---or by changing the average topological charge---(a) to (c). The control over the topological charges is experimentally achieved without any physical adjustments to the light spring generating setup (d), as it relies on changing the spatiospectral hologram (e) displayed on the spatial light modulator. The experimental demonstration of the $v_{og}$ tunability from sub-(g) to superluminal (h) can be seen in (f). The white squares represent the discrete $v_{og}$ values programmed through the holograms, this represents the expected $v_{og}$ values for the experimentally extracted light spring radius, corresponding to Eq. \ref{eqOGV2}. The black circles are the values retrieved experimentally from the tracking of the light spring hotspot angle rotation, while the pulse propagates as the experimentally extracted light spring radius, corresponding to Eq. \ref{eqOGV1}. The continuous line is a numerical sweep for the same bandwidth and fixed spatiospectral slope, interpolating between the discrete experimental settings. Panels (g) and (h) show reconstructed light springs with subluminal and superluminal $v_{og}$, respectively. $\mathcal{}f$ represents the laser frequency, and $\ell$ topological charge. The scale bar in (g) and (h) is 345 $\mu$m, and $\mathcal{}f$ represents the laser frequency.}
\end{figure*}

Whereas these space-time correlations can result from misalignments in chromatic optics in the laser chain, synthesizing tunable space-time beams requires to realize the desired correlations between the spatial and temporal/frequency components in a coherent light pulse \cite{STwavepackets}. For this purpose, the use of advanced optical elements, such as flat optics, metasurfaces~\cite{MS,MS2} and spatial light modulators (SLMs)~\cite{Rosales_SLM,SynthesisSheets}, is playing a crucial role in the development of space-time beams and their applications. A new type of space-time beam synthesized thanks to these technologies is the light spring (LS). A LS has a helical wavepacket \cite{QuereLS,MarcoLS,LS2,LS3}, which means that an ultrashort pulse is structured to create a spatiotemporal spring-like structure. This helical wavepacket results into an orbiting intensity profile when considering its intersection with a plane orthogonal to the LS propagation direction. Previous works have established the synthesis of light springs and the control of their topological-spectral correlations, and identified the spatiospectral slope $\Omega=\partial\omega/\partial\ell$ as a descriptor of their rotation dynamics~\cite{QuereLS,MarcoLS,LS3}. Here we focus on the tangential speed of the rotating intensity maximum itself, which is the quantity directly relevant for orbital motion and for comparison with $c$.

In this work we introduce the orbital group velocity, $v_{og}$, i.e., the velocity at which the peak intensity of a LS rotates, as a novel tunable degree of freedom of these spatiotemporal beams. We experimentally demonstrate its tunability to values below and above $c$, and use that control to identify an application regime in which a superluminal $v_{og}$ can drive plasma superradiance. The optical result is the primary experimental advance, while the plasma simulation study shows how the same control parameter can be leveraged in a high-energy interaction scenario.

Light springs are characterized by a linear correlation between the spectral components of an ultrafast laser pulse and the topological charge of their corresponding spatial modes. At a given plane perpendicular to the propagation of the laser pulse, a LS results in an orbiting intensity distribution around the propagation axis. For an orbital angular momentum (OAM) mode with phase $\Phi = kz + \ell\phi$, the azimuthal component of the local wavevector is $k_{\phi}=r^{-1}\partial\Phi/\partial\phi=\ell/r$. Evaluated at the hotspot radius $r=r_{LS}$, the orbital analogue of a group velocity is therefore
\begin{eqnarray}
v_{og} = \frac{\partial{\omega}}{\partial{k_{\phi}}} = \frac{\partial{\omega}}{\partial{\ell}}r_{LS}\label{eqOGV2},
\end{eqnarray}
where $\omega = 2\pi f$ and $\ell$ are the spectral angular frequency and topological charge of the LS (see Suppl. Sec. IA). Equation \ref{eqOGV2} includes the slope of the spatiospectral correlation, and multiplication by $r_{LS}$ yields the tangential speed of the peak intensity. While $\Omega$ has been reported previously \cite{QuereLS, LS3}, $v_{og}$---naturally compared to $c$---provides the appropriate figure of merit to quantify orbital motion and to identify new interaction regimes. Similarly to the well-known, widely tunable $v_{g}$ of space-time beams \cite{AbouraddyGV}, this orbital motion can be tailored to achieve sub- and superluminal values. Therefore, such a parameter should be common to all helical wavepackets, but its proper description and experimental demonstration has been lacking.

%%% Figure: Plasma interaction (1/2)
\begin{figure*}[t] 
\includegraphics[scale=1]{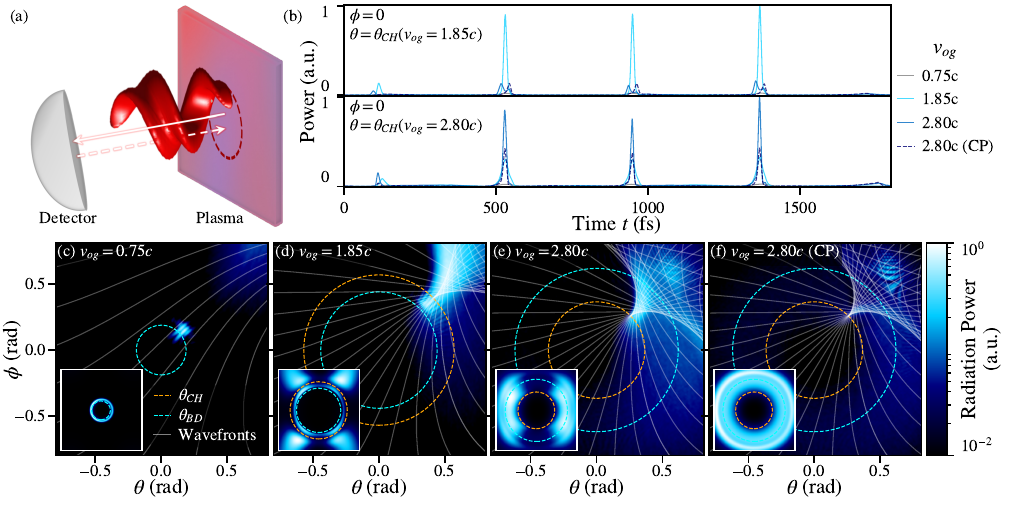}
\caption{\label{fig:plasma_radiation} \textbf{Superluminal light springs unlocking plasma superradiance in PIC simulations.} A LS reaches a thin overdense plasma, while a detector is placed collinearly to the LS to measure the radiation produced from the laser-plasma interaction (a). The detector is spherical and captures the radiation over time. Plot (b) shows the intensity of the radiation captured in each simulation at two positions in the detector, corresponding to the Cherenkov angles for velocities $v_{og}=1.85c$ (top) and $v_{og}=2.80c$ (bottom). Plots (c)-(f) show the instantaneous power at $t=473$ fs, while the insets show the total energy captured throughout the simulations. The angles $\phi$ and $\theta$ represent the angle with respect to the z-y plane and the z-x plane, respectively. $\theta_{CH}$ and $\theta_{BD}$ are the Cherenkov and beam divergence angles, respectively.}
\end{figure*}
%%%%%

An interesting analogy of a LS can be found in the light emitted by a lighthouse (Fig. \ref{fig:Concept}). We use this analogy in a strictly kinematic sense. In a lighthouse, the beam rotates with angular velocity $\dot \alpha = d\alpha/dt$ producing a moving spot at radius $r_{LH}$ with tangential speed $v_{LH} = \dot \alpha ~r_{LH}$. For sufficiently large $r_{LH}$,  $v_{LH}$ can exceed $c$. This does not imply superluminal transport of energy or information as the spot is not a material entity, and the photons themselves propagate at light speed. Rather, successive photons illuminate successive positions. A LS does not originate from a rotating source. Instead, its rotating hotspot arises from the spatiotemporal interference of spectral components carrying different topological charges. Nevertheless, on a fixed transverse plane, both systems share the same apparent orbital kinematics. Accordingly, for a LS, on a transverse plane, its intensity maximum rotates with angular position $\alpha(t)$ obeying
\begin{eqnarray}
v_{og} = \dot \alpha ~r_{LS}\label{eqOGV1}.
\end{eqnarray}
Thus, for a LS, Eq. \ref{eqOGV1} is equivalent to Eq. \ref{eqOGV2}, and $v_{og}$ quantifies the apparent motion \cite{SyntMotion1,SyntMotion2,SyntMotion3} of the maximum intensity of the wavepacket, while no photon travels above $c$ (see Suppl. Sec. IB).
  
%%% OGV %%%
Synthesizing a LS with an ultrashort laser pulse requires controlling the topological charge of different spectral components of a laser pulse with OAM. The experimental setup used to demonstrate the control over $v_{og}$ can be seen in Fig. \ref{fig:OGV}d. A titanium-sapphire laser (40 fs pulses at 1 kHz repetition rate from VOXEL high-intensity laser facility, see Suppl. Sec III) with a bandwidth of 13 THz was used to generate the LS. The ultrashort pulses reach an axicon grating (Suppl. Sec. II) \cite{MarcoLS}, which diffracts different spectral components at different angles. Using a lens to conjugate the far field of the axicon, we generate a ring with a radially varying frequency distribution, where a reflective SLM is then used to impart an azimuthal phase with a radially varying topological charge, and thus generate the desired spatiospectral correlation for the LS (Fig.~\ref{fig:OGV}e). The SLM is the main component that limits the energy sustained by the system. Nonetheless, it was used for the reconfigurability it adds to the system, and simplifies the $v_{og}$ tunability implementation. Additionally, for applications requiring LS with higher pulse energies, the hologram displayed on the SLM can be replaced by a fused silica hologram (see Suppl. Sec IV). The laser pulse with the imparted spatiospectral coupling reflects off the SLM and is focused by the same conjugation lens to form the desired helical intensity distribution at the focal plane. As shown in Fig.~\ref{fig:OGV}d, a thin beamsplitter is used to separate the incoming laser pulse from the LS before reaching the CMOS detector. This system is also known as a Fourier synthesizer, where the frequencies are spatially separated, and the OAM content is introduced via phase modulations.

Once the LS is synthesized, we reconstruct its intensity at the Fourier plane of the lens by measuring the interference of the LS with a reference pulse at this location. This reference pulse is taken from the incoming laser pulse with a beamsplitter placed prior to the axicon, after which it traverses an optical delay line to control its relative delay with respect to the LS at the measuring plane, where the detector is placed (Fig.~\ref{fig:OGV}d). Varying the relative delay between the LS and the reference pulse at the detector plane, we can reconstruct the temporal evolution of the intensity distribution of the LS at this location~\cite{AbouraddyGV}. This approach also allows us to characterize LS with different tunable parameters, such as different topological charges or a number of independent spectral radii (Suppl. Sec. III).

As prescribed by Eqs. \ref{eqOGV2} and \ref{eqOGV1}, there are two options to experimentally tune the $v_{og}$ of a LS: either by tuning the slope of the spatiospectral correlation or the radius of the LS. Figure~\ref{fig:OGV}a-c illustrates how to increase the $v_{og}$ from $0.6c$ (Figure~\ref{fig:OGV}a) to $1.1c$ by either changing $\Omega$ (Figure~\ref{fig:OGV}b) or by enlarging the radius via an increase of the average topological charge (Figure~\ref{fig:OGV}c).  In principle, changing any of these variables will give access to sub- and superluminal regimes of $v_{og}$ values. However, experimentally, tuning $v_{og}$ by controlling the slope of the spatiospectral correlation limits the $v_{og}$ range. The slope of the spatiospectral correlation can be tuned by changing the frequency bandwidth used or the total variation of topological charge applied to create the LS (Eq. \ref{eqOGV2}). Once a given experimental setup is built, the frequency bandwidth achieved by it is determined by the laser characteristics: decreasing it is often detrimental, as it reduces the degrees of freedom in the Fourier synthesis of the space-time beam, while increasing it typically requires complicated experimental setups that rely on nonlinear optics \cite{Attosecond}. Alternatively, the control over the $r_{LS}$ is simpler to implement experimentally, and also offers broader access to $v_{og}$ regimes. This can be realized by either using a magnifying optical relay, or by changing the average topological charge of the LS~\cite{RadiusOAM}. The latter is the method used in this work to tune the $v_{og}$ of the synthesized LS from subluminal to superluminal values, as it only requires changing the hologram applied at the SLM without any physical alterations to the experimental setup. This distinction is important relative to earlier light-spring demonstrations as the present experiment is not only synthesizing a helical wavepacket, but directly sweeping the hotspot speed across interaction regimes.

The $v_{og}$ is set by the spatiospectral correlation used to synthesize the LS (Eq. \ref{eqOGV2}) and is therefore programmed through the hologram displayed on the SLM. To verify that the programmed orbital group velocity matches the actual $v_{og}$, we independently extract it by tracking the azimuthal angle of the intensity hotspot as the LS propagates, the black dots in Fig. \ref{fig:OGV}d (Eq. \ref{eqOGV1} and Suppl. Sec. V), and compare it to the programmed $v_{og}$, i.e., the value applied when creating the hologram for the same experimentally extracted radius, as the white squares (Eq. \ref{eqOGV2}). We also perform numerical simulations of the LS to benchmark the $v_{og}$ tunability. The simulated $v_{og}$ are derived numerically, taking into account a Gaussian beam profile with a fixed waist and the same frequency bandwidth of the experimental laser pulse. Then the spatiospectral correlation is imposed, and the resulting electromagnetic field is calculated assuming a superposition of Bessel-Gauss modes, created by an axicon grating (Suppl. Sec. VI). 

Fig. \ref{fig:OGV}f exhibits a very good agreement between the simulated, programmed, and experimentally measured $v_{og}$. The programmed $v_{og}$ relies on assuming that the axicon and SLM are creating the correlation that is given between the frequency bandwidth and the applied topological charges, which cannot be directly characterized and is only inferred by experimental setup characterization. On the other hand, the experimentally measured $v_{og}$ is directly extracted from the LS intensity profile, depending only on the accuracy of the retrieval method. Even considering the inherent experimental errors, mainly coming from the pixel-size limitation imposed by the SLM and aberrations added in the optical path, it agrees well with the simulated and programmed $v_{og}$ values, highlighting the robustness of this method in generating and controlling the LS properties, particularly the $v_{og}$. This shows that changing the topological charges used to create the LS, the $v_{og}$ can be tuned from subluminal (Fig. \ref{fig:OGV}g) to superluminal (Fig. \ref{fig:OGV}h) values, demonstrating the wide tunability of the $v_{og}$.

%%% LS interaction %%%
The sub- and superluminal regimes of $v_{og}$ unlock new control over light-matter interactions. Subluminal velocities are important when considering solid-state interactions \cite{SolidState} or laser-plasma accelerators \cite{JorgePlasmaAcceleration,VieiraHighOrbitalAngular2016}. In these scenarios, coupling the $v_{og}$ with gas jets or underdense plasmas can excite spatiotemporal structured plasma waves and control their motion, while enhancing particles' energy gain and trapping efficiency. Here, instead, we focus in superluminal velocities and show by particle-in-cell simulations (Osiris \cite{OSIRIS}; Suppl. Sec. VII) and radiation diagnostics (RaDiO \cite{RaDiO}), that in an overdense plasma \cite{LichtersShortpulseLaserHarmonics1996, DenoeudInteractionUltraintenseLaser2017, LeblancPlasmaHologramsUltrahighintensity2017} the LS allows the control of the emitted radiation intensity, unlocking a superradiant effect based on superluminal~\cite{Superradiance2} quasiparticles~\cite{SuperradianceQuasiparticle}. This method is then another way of generating THz radiation from plasma interactions~\cite{THzFlyingFocus}, in par with, for example, two color laser pulses techniques ~\cite{THzTwoColor, THzCherenkov}. For computational efficiency, the application study models the driver as a superposition of Laguerre-Gaussian modes chosen to reproduce the desired $v_{og}$; the underlying mechanism depends on the orbital motion of the hotspot rather than on the exact modal basis.

%%% Table: Simulation parameters
\begin{table}[b]
\caption{\label{tab:SimMain} Main parameters of the four LS used in the particle-in-cell simulations. All cases use $\lambda_0=1~\mu$m, $w_0=5~\mu$m, $\tau \approx 1.67$ ps, and total energy $0.5$ J. We consider linear (LP) and circular (CP) polarizations.}
\begin{ruledtabular}
\begin{tabular}{cccccc}
$v_{og}/c$ & $\ell_0$ & $r_{LS}$ ($\mu$m) & $\theta_{CH}$ (º) & $\theta_{BD}$ (º) & Pol. \\
0.75 &  18 & 15.00 &     - & 10.82 & LP \\
1.85 & 110 & 37.08 & 32.64 & 25.28 & LP \\
2.80 & 250 & 55.90 & 20.96 & 35.46 & LP \\
2.80 & 250 & 55.90 & 20.96 & 35.46 & CP \\
\end{tabular}
\end{ruledtabular}
\end{table}
%%%%%

%%% Figure: Plasma interaction (2/2)
\begin{figure}[hb!] 
\includegraphics[scale=1]{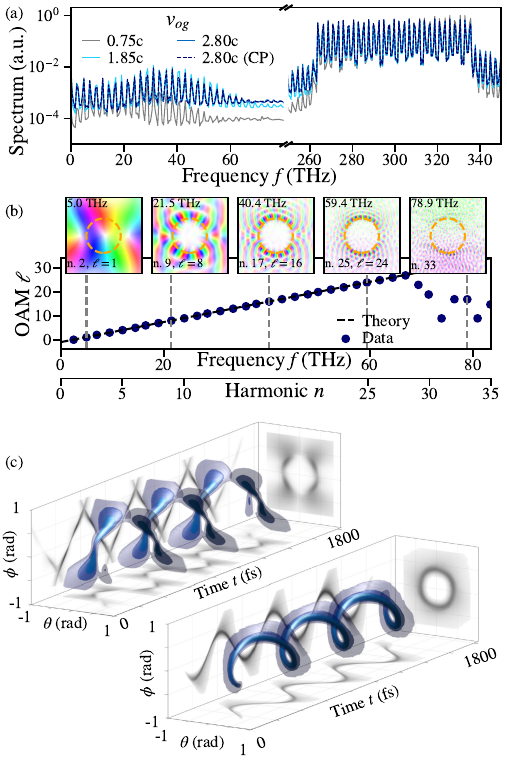}
\caption{\label{fig:plasma_spectrum} \textbf{Spatiospectral analysis of the structured terahertz radiation.} The spectra of the radiation captured by the detector in each simulation can be seen in (a). We focus on the drivers' high frequency range and the low frequency domain, responsible for the superradiant radiation. Plot (b) displays the topological charge of each of the THz harmonics for the linear polarization $v_{og}=2.80c$ case, with the insets showing examples of spectral phase for five different harmonics. Each harmonic has a different topological charge, characteristic of a LS (see Suppl. Sec. VIII D for remaining cases). Isosurface plots of the intensity of the generated THz LS with $v_{og}=2.80c$ are seen in (c), for linear (top) and circular (bottom) polarization drivers respectively. The circular polarization creates a cylindrical symmetric radiated LS. }
\end{figure}
%%%%%

A quasiparticle is a localized nearly-constant current-density profile. Because it is a result of a collective effect, and unlike point-like particles (e.g., electrons), quasiparticles can execute arbitrary trajectories and travel at arbitrary speeds (even superluminal). The coherent part of the quasiparticle radiation spectrum is identical to that emitted by a single finite-sized particle undergoing a similar motion. The flexibility to control the quasiparticle motion opens new radiation mechanisms that are usually forbidden to single charged particles. For example, a quasiparticle traveling faster than $c$ leads to quasiparticle Cherenkov radiation~\cite{Superradiance2, SuperradianceQuasiparticle, pardal_SuperradiantScatteringEvanescent2026}. Here, we propose to use a LS to unlock a novel superradiant effect based on a superluminal quasiparticle, traveling along a synchrotron (i.e. circular) trajectory in laser-matter interactions. The effect is due to the onset of an optical shock, which appears as a result of constructive interference of the radiation emitted by the quasiparticle during its trajectory, and at the Cherenkov angle. In this geometry, the Cherenkov angle is given by $\theta_{CH} = \sin^{-1}(c/v_{og})$ measured with respect to the axis of rotation. This definition of the Cherenkov angle differs from the standard $\cos^{-1}(c/v)$ as the geometry we tackle here is different.

Fig.~\ref{fig:plasma_radiation}~a portrays a schematic of our simulated setup. The parameters of the used LS are shown in Table~\ref{tab:SimMain}: these were chosen to closely resemble the experimental tunability demonstrated before by controlling the $v_{og}$ with the average topological charge and using linear polarized drivers. We also tested a circular polarized LS to assess the effects on the generated radiation. All the lasers used were composed of 31 modes with frequencies uniformly distributed between 263.86 and 335.43 THz, yielding an average frequency of 300 THz with a spacing of 2.39 THz. The LS travels along the $z$ direction and shine on a cold overdense plasma slab of density $n_0=2.824\times10^{21}{\rm cm}^{-3}$. A virtual spherical detector is placed in the negative z-direction to detect the far field of the radiation emitted by the plasma electrons when interacting with the laser fields \cite{RaDiO}. Fig.~\ref{fig:plasma_radiation}~b displays the radiation power over time captured by the detector for each simulation at two positions, corresponding to the Cherenkov angles for velocities $v_{og}=1.85c$ (top) and $v_{og}=2.80c$ (bottom). In each of these plots, the radiation from the simulation with the LS with the corresponding $v_{og}$ resembles a train of unipolar pulses spaced by the LS period of rotation, i.e.,  around 420 fs (Suppl. Sec. VIII A). Figs.~\ref{fig:plasma_radiation}~c-f show the instantaneous power captured by the detector in each simulation. The insets show the total integrated radiation intensity throughout the complete interaction. All regimes include the pure reflection of the driver LS at the beam divergence angle $\theta_{BD} = \frac{\sqrt{\ell_0}\lambda_0}{\sqrt{2}\pi w_0}$, where $w_0$, $\ell_0$, and $\omega_0$ are, respectively, the LS spot size, average topological charge, and main frequency (see Suppl. Sec. VII for more details). However, for the superluminal cases (Figs.~\ref{fig:plasma_radiation}d-f), an extra intensity feature is present at the Cherenkov angle, corresponding to the cusp of a cardioid shaped optical shock. This shock arises from the intersection of the instantaneous wavefronts, which are plotted as white lines in the figures. A detailed description of these wavefronts is included in the Supplementary Sec. VIII B. It is clear that the angle at which the cusp appears is controlled by simply changing the superluminal velocity of the driver, once again highlighting the need for an accurate tunability of the $v_{og}$. Regarding polarization, when using a linearly polarized LS driver, the resulting radiation has intensity minima at $x=0$, while a circularly polarized driver generates purely axisymmetric radiation. This asymmetry is explained by the shape factor of the quasiparticle radiation \cite{SuperradianceQuasiparticle} (see Suppl. Sec. VIII C for more details).

For a more comprehensive understanding of the radiation generated, the data was spatially and spectrally analyzed and the results shown in Fig.~\ref{fig:plasma_spectrum}. The spectrum over the relevant frequency ranges in shown in Fig.~\ref{fig:plasma_spectrum}~a. The pure reflection of the drivers is seen between 263.86 and 335.43 THz as the 31 most intense peaks. On the other end of the spectrum, in the low frequency range between 2.39 and 71.57 THz, the superradiant content is shown. It is composed of well defined peaks at harmonics of the frequency associated with the temporal pitch of the LS, i.e, 2.39 THZ. While for the subluminal case, the intensity quickly falls after a few harmonics, for the superluminal cases, the intensity of the harmonics is much higher and thus the harmonic content much broader. It is important to note that these harmonics are phase locked, thus effectively constituting a frequency comb with phase coherence inherited from the infrared LS laser drive, and have an azimuthal index corresponding to the harmonic order. These characteristics are a direct consequence of the ultrafast spatiotemporal dynamics of the LS used. This can be seen in Fig.~\ref{fig:plasma_spectrum}~b where the spatial phase distribution is plotted for different THz harmonics. Each harmonic has a different topological charge, just as the driver. Therefore, the THz radiation generated is itself a LS. Figs.~\ref{fig:plasma_spectrum}~c and d show isosurface plots of the spectrally filtered radiation on the detector, considering only the THz range. As seen before, the linear polarized case yields a cylindrically asymmetric THz LS while the circular polarization one results in a much cleaner and symmetric LS.

Regarding the efficiency of the process, the energy of this train of THz pulses is $\approx0.5\%$ of the energy of the LS drive in the superluminal regimes (see Suppl. Sec. VIII E. for more details), which is on par with \cite{jang_EfficientTerahertzBrunel2019, tailliez_TerahertzPulseGeneration2020}. This configuration thus yields a way of generating pulsed radiation is in the terahertz range - a strategic region of the electromagnetic spectrum - and in the millijoule level, unlocking applications in structured strong-field terahertz science \cite{pak2023multi,liao2019multimillijoule}.

%%% Discussion%%%
Therefore, the proper tuning of the $v_{og}$ of the LS promises unprecedented control over laser-matter interactions at each $v_{og}$ regime accessing new interaction scenarios, particularly suited for the acceleration of plasma to generate relativistic electron vortex beams \cite{JorgePlasmaAcceleration} or as a tunable terahertz source of radiation, based on the Cherenkov mechanism. To experimentally measure such effects, it will be necessary to generate high-energy LS. Since the setup currently relies on a SLM, the energies that can be used to generate the LS are limited by it, but this problem could be circumvented by high-damage-threshold holograms fabricated in silica \cite{HighPower}.

Additionally, when creating a LS, many parameters can be tuned, depending on how the axicon and holograms are generated. This opens the possibility to create LS with a radial gradient of $v_{og}$, precisely optimized to deliver different $v_{og}$ at different radial positions, or by implementing the control of the LS chirp, to create a different frequency response based on the propagation of the LS. Lastly, as shown in this work, the generation of tunable space-time beams is currently at the frontier of unlocking new sources of radiation, while holding the promise to be the next generation of ultrafast light sources in strategic regions of the electromagnetic spectrum.

\section*{Acknowledgments}
The authors would like to thank Slava Smartsev and Jerome Faure for their involvement in the discussion of the techniques for light spring characterization.
G.V. and M.P. received funding from the European Research Council (ERC StG) under the European Union’s Horizon Europe research and innovation program (Grant agreement No. 101161858). M.P. wishes to acknowledge funding of the Research Unit INESC MN from the Fundação para a Ciência e a Tecnologia (FCT) through Plurianual financing (UIDB/05367/2025, UID/PRR/5367/2025 and UID/PRR2/05367/2025; DOI: 10.54499/UIS/PRR/05367/2025, 10.54499/UID/PRR2/05367/2025). The work of R.A. is supported by Fundação para a Ciência e Tecnologia (FCT, Portugal) grant number UI/BD/154677/2022 (DOI: 10.54499/UI/BD/154677/2022). This work was supported by FCT I.P. under Project 2024.07895.CPCA.A3 (DOI: 10.54499/2024.06987.CPCA.A3) at MareNostrum 5 supercomputer, jointly funded by EuroHPC JU, Portugal, Turkey and Spain; under Project 2023.14906.PEX (DOI: 10.54499/2023.14906.PEX); and under Project EHPC-EXT-2025E01-103.
PC, RN, and MF were funded by European Council Pathfinder Project 101047223, and LaserLab-Europe (grant agreement no. 871124). 

%%% Citations %%%
%

%%%%%%%%%%%%%%%%%% Supplementary Information %%%%%%%%%%%%%%%%%%%
\setcounter{figure}{0}
\setcounter{table}{0}
\setcounter{equation}{0}
\widetext

%%%%% LH - LS
\section*{Supplementary Material for:\\
Space-time beams with tunable orbital group velocity toward plasma superradiance}

\section{Orbital group velocity}

In this work, the orbital group velocity ($v_{og}$) is introduced as a new tunable parameter for helical wavepackets, and the lighthouse comparison is explicitly framed as a kinematic analogy. In this section, we provide an expanded derivation of $v_{og}$ and further clarify the limits of that analogy.

\subsection{Definition of orbital group velocity}

In cylindrical coordinates, an orbital angular momentum (OAM) mode can be written as
\begin{eqnarray}
    E(r,\phi,z) \sim A(r,z)e^{i\Phi(r,\phi,z)}, \Phi = kz + \ell\phi.
\nonumber
\end{eqnarray}
The local wavevector is
\begin{eqnarray}
    \textbf{k}(r) = \nabla\Phi.
\nonumber
\end{eqnarray}
This becomes
\begin{eqnarray}
    \nabla\Phi = \textbf{ê}_r\frac{\partial\Phi}{\partial r} + \textbf{ê}_{\phi} \frac{1}{r} \frac{\partial\Phi}{\partial \phi} + \textbf{ê}_z\frac{\partial\Phi}{\partial z}.
\nonumber
\end{eqnarray}
The azimuthal component is
\begin{eqnarray}
    k_\phi(r) = \frac{1}{r}\frac{\partial\Phi}{\partial\phi} = \frac{1}{r}\frac{\partial(\ell\phi)}{\partial\phi} = \frac{\ell}{r}.
\nonumber
\end{eqnarray}
The group velocity is typically written as $v_g=\frac{\partial\omega}{\partial k}$, thus we define its orbital counterpart in the azimuthal direction and at the hotspot radius $r=r_{LS}$ using the chain rule
\begin{eqnarray}
    \frac{\partial \omega}{\partial k_\phi} = \frac{\partial \omega}{\partial \ell} \cdot \frac{\partial \ell}{\partial k_\phi} = \frac{\partial \omega}{\partial \ell} \cdot r_{LS} = v_{og}.
\nonumber
\end{eqnarray}

\subsection{Light spring and lighthouse relation}
The lighthouse effect is a phenomenon that describes the apparent superluminal motion of the light source in a lighthouse. The lighthouse has a light source that is rotated with an angular velocity of $\frac{\partial{\alpha}(t)}{\partial{t}}$, which produces a moving spot at a given radius $r_{LH}$ with linear velocity,

\begin{eqnarray}
v_{LH}=\frac{\partial{\alpha}(t)}{\partial{t}}r_{LH}.
\end{eqnarray}

It can be easily seen that $v_{LH}$ can assume arbitrarily large values. Particularly at a distant $r_{LH}$, it can exceed the velocity of light without breaking relativistic causality. For this case, the effect is a result of different locations being illuminated by different photons emitted at different times by the light source. On the other hand, in a given fixed orthogonal plane z, the light spring (LS) intensity hotspot describes a circular motion with linear velocity,

\begin{eqnarray}
v_{og}=\frac{\partial{\alpha}(t)}{\partial{t}}r_{LS}\label{EqOGV1}\\
      =\frac{\partial{\omega}}{\partial{\ell}}r_{LS}\label{EqOGV2}, 
\end{eqnarray} where $r_{LS}$, $\omega$, and $\ell$ are the ls radius, angular frequency and topological charge, respectively. $\alpha$ is parameterized by the laboratory time on a fixed transverse plane. There is no break in relativistic causality. The superluminal motion observed arises from the angular arrival time of the photon, no energy or information is being transported. Therefore, the lighthouse spot and the LS intensity hotspot can be understood in the same way: in both cases one is observing an azimuthally evolving light pattern whose apparent transverse motion is set by how fast the pattern angle changes and by the radius at which it is observed. In this sense, the two systems share the same kinematic description of a rotating intensity pattern, even though the underlying physical mechanisms that generate the rotation are different.

%%%%% Experimental setup
\section{Axicon design and specifications}

The axicon used in this work is a multilevel axicon grating designed to implement, over each radial period, an approximate $0$ to $2\pi$ phase ramp at the central wavelength of the titanium-sapphire source. 
The axicon has diameter $D = 10~\mathrm{mm}$ and radial period $d = 10~\mu\mathrm{m}$. Each period is quantized into $N=10$ phase levels. A master is fabricated by grayscale lithography, then the replica is fabricated in a UV-cured hybrid polymer (OrmoComp) using UV nanoimprint lithography. Ormocomp has a refractive index at the titanium-sapphire central wavelength of $1.51$ and robustness similar to that of glass, making it suitable for operation with high-energy femtosecond pulses. A Gaussian beam with wavelength $\lambda$ impinging on the axicon forms in the Fourier plane (reached using a lens with focal length $f$) a beam with annular radius given by
\begin{equation}
R(\lambda) \simeq f\,\frac{\lambda}{d}.
\label{eq_axi_R}
\end{equation}
The beam waist $w_0$ of the input Gaussian sets the radial half-thickness of the focused annulus as
\begin{equation}
T(\lambda) \simeq \frac{f\lambda}{\pi w_0}.
\label{eq_axi_T}
\end{equation}
The criterion for the number of independent spectral points in the design of an $\ell(f)$ correlation is that neighboring annuli do not overlap. Given our beam waist of $w_0\sim2.5$ mm and focal length $f=75$ mm, we have around 10 independent spectral windows within the bandwith of the titanium-sapphire laser for the synthesis of light spring' spatiospectral correlations.

\section{Characterization of experimental variables to generate light springs}

The experiments were carried out in the VOXEL high-intensity laser facility at Instituto Superior Técnico, and as the light source, a titanium-sapphire laser (central wavelength of 800 nm), with nominal pulse energy of 3 mJ, beam-size at the output of 5 mm, and horizontal polarization.

\begin{figure*}[ht]
\includegraphics[scale=0.5]{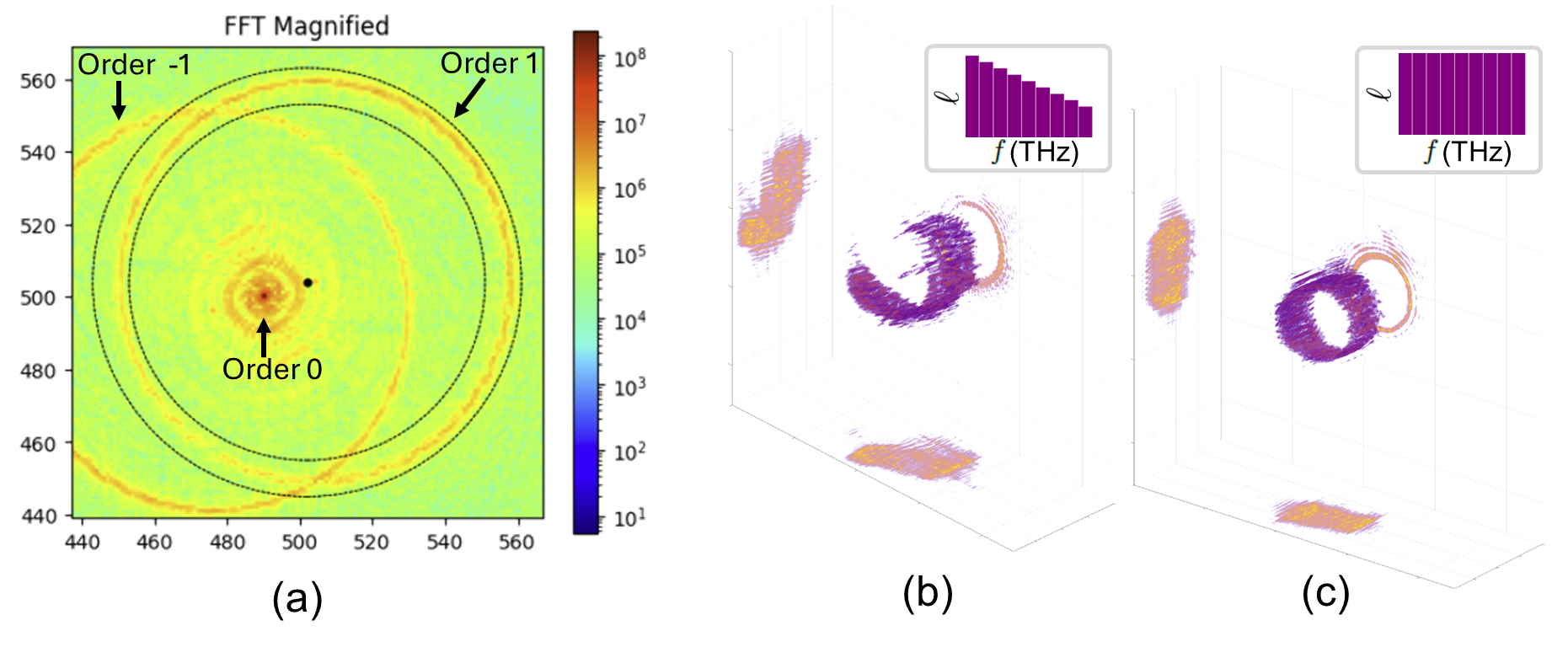}
\caption{\label{fig:FFT} \textbf{Reconstruction of the LS isosurface.} The angle between the arms of the interferometer is decisive to the proper reconstruction of the LS. It directly impacts the FFT (a), which is used to reconstruct the intensity profile of the LS, used to plot the LS isosurface, (b) and (c). (b) Shows a LS isosurface with linear space-time correlation, and (c) shows a torus isosurface, a structure where the same topological charge is applied to all frequencies to create the spatiospectral correlation. In (b) and (c) $\ell$ represents the topological charges and $\mathcal{}f$ the laser frequencies.}
\end{figure*}

To generate and reconstruct the LS, there are experimental variables that need to be optimized individually. Once the LS is synthesized, it is important to optimize the reconstruction. The technique used to reconstruct the LS in this work is off-axis digital holography (OADH) \cite{AbouraddyGV_SI}. The most important parameter is properly adjusting the angle between the interferometer arm that has the LS and the reference arm. When using OADH, the amplitude and phase of the electromagnetic field are encoded on the first order of the fast Fourier transform (FFT) (Fig. \ref{fig:FFT}a). Typically, the angle between the arms is maximized, so that the zero order does not add noise to the reconstruction. However, when reconstructing a LS, a bigger angle means a bigger tilt in the LS isosurface. To avoid this, the angle is set as small as possible, with the zero order of the FFT lying between the orders 1 and -1 (Fig. \ref{fig:FFT}a).  

\begin{figure*}[h!]
\includegraphics[scale=0.5]{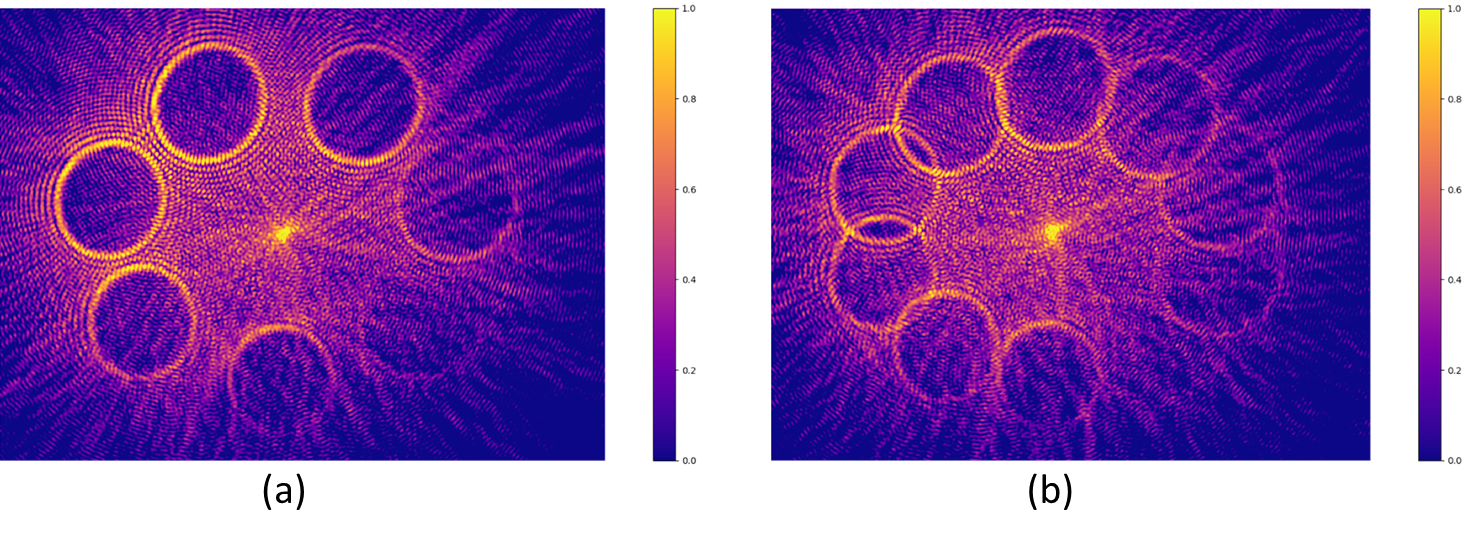}
\caption{\label{fig:LaserFlowers} \textbf{Experimental calibration of the axicon and SLM.} To test the control of the spectral windows created with the axicon grating, in addition to the topological charges applied to each window, a linear diffraction grating was added to the hologram displayed at the SLM. This allows to quantify the control over the spectral independent points (7 (a) and 10 (b)) with the SLM.}
\end{figure*}

Once the reconstruction of the LS is optimized (Fig. \ref{fig:FFT}b and c), it is important to analyze how the axicon and SLM are creating the LS. The axicon is designed to separate the frequencies of the laser in a first-order beam, with a radially-dispersed spectrum. The spatial light modulator (SLM) is used to apply a topological charge to each one of the radii. To guarantee that the axicon and the SLM are giving the beam the desired properties, in addition to the topological charges, for the purpose of the characterization discussed in this section the hologram encodes an additional linear diffraction grating. The linear diffraction grating is applied such that each considered spectral window for the spatiospectral correlation is diffracted at a different spatial position. This way, at the CCD, it is possible to see spatially separated rings (Fig \ref{fig:LaserFlowers}). This allows to quantify the number of independent experimental spectral windows the Fourier synthesizer can control SLM.

The axicon was designed to guarantee around 10 independent spectral windows (Sec. I): this number is calculated considering a given beam waist and central wavelength, and ideal blaze profile and optical components. The tests reported in Fig. \ref{fig:LaserFlowers}b show that it is indeed possible to resolve at least 10 different circles by the procedure described above, encoding a linear blazed grating in the hologram. We expect that by adjusting the beam waist it should be possible to push the effective number of spectral windows above 10, even for a narrowband source like the titanium-sapphire laser used in this work.

\section{Analysis of the Fourier synthesizer efficiency and scalability to high pulse energies}

The experimental setup used in this work was designed as a flexible and fully reconfigurable Fourier synthesizer for the demonstration and characterization of LS with tunable orbital group velocity, rather than as a high-energy source architecture. In the present implementation, a Ti:sapphire laser is sent to an axicon grating and the spatiospectral phase correlation is programmed with a reflective SLM placed at the Fourier plane of the axicon. This choice is ideal for proof-of-principle experiments because it allows us to rapidly change the topological-spectral correlation and directly explore the subluminal and superluminal regimes of $v_{og}$. However, the SLM is also the element that ultimately limits operation at high pulse energy. In addition, the present setup includes beamsplitters and a reference arm because the LS is characterized by off-axis digital holography. These components are required for metrology, but they are not part of the minimum architecture needed to generate the beam itself. In a source-oriented implementation aimed at laser-plasma experiments, the diagnostic interferometer can be removed increasing throughput and removing low-damage or unnecessary elements from the high-energy beam path.

The efficiencies of the axicon and the SLM used in this work are approximately 71\% and 97\%, respectively. Therefore, for the laser used here with nominal pulse energy of 3 mJ, the maximum pulse energy output of the setup is 2.1 mJ. This is still 2 orders of magnitude lower than the pulse energy of $0.5$~J used in our PIC simulations (Sec. \ref{sec:PIC}) to reveal the superradiant mechanism. However, our synthesis principle itself is not tied to the use of an SLM. The generation of the LS relies only on two passive linear operations: first, angular spectral dispersion performed by the axicon grating and second, application of a prescribed azimuthal phase profile in the Fourier plane. Therefore, once the desired correlation has been identified, the reconfigurable SLM can in principle be replaced by a static high-damage-threshold diffractive element that implements the same hologram. In that case, the complete LS synthesizer becomes a passive Fourier optical system composed only of high-damage-threshold optics. This is the natural route towards Joule-class operation.

The relevant scaling is governed by fluence rather than by pulse energy alone. In other words, increasing the pulse energy does not by itself invalidate the architecture, provided that the illuminated area on the passive optics is increased accordingly so that the local fluence remains below the damage threshold. Fused silica is one of the most used substrates for high-energy applications, and for the laser considered in this work has a damage threshold of the order of J/cm$^2$. Moreover, scaling is particularly favorable in the present geometry because the spectral content is not focused to a compact spot at the shaping plane, but distributed over an annulus. For an axicon with period $d$ and a Fourier lens of focal length $f$, a wavelength $\lambda$ is mapped to an annular radius $R$ given by Eq. \ref{eq_axi_R} with radial half-thickness $T$ given by Eq. \ref{eq_axi_T}. Hence, the energy incident on the Fourier-plane hologram is naturally spread over an annular area, and this area can be further increased by a suitable choice of beam size and relay optics, while preserving the same programmed $\ell(\omega)$ correlation. Therefore, the combination of an all-silica axicon, already used in this work, and all-silica holograms will enable scaling to Joule level for laser-plasma superradiance based on $v_{og}$, while maintaining the system efficiency and structure quality.

\section{Experimental angle extraction of light spring intensity profile}

To experimentally measure $v_{og}$, two quantities need to be extracted: the angle variation of the LS hotspot ($\frac{\partial\alpha(t)}{\partial{t}}$) and the LS radius ($r_{LS}$) (Eq. \ref{EqOGV1}).

\begin{figure*}[ht]
\includegraphics[scale=0.56]{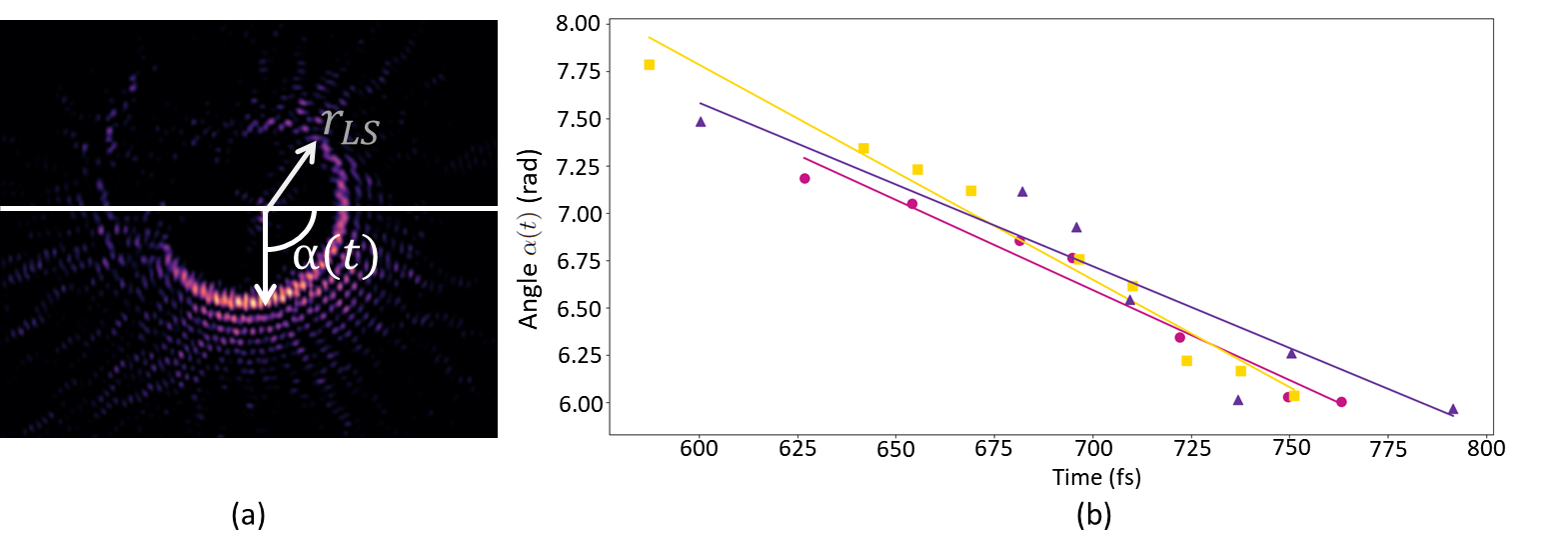}
\caption{\label{fig:AngleExt} \textbf{Experimental measurement of the intensity hotspot rotation as the LS propagates}. To experimentally measure the $v_{og}$, the angle ($\alpha(t)$) variation of the LS intensity hotspot (a) is tracked as the helical wave packet propagates (b). In (b), the pink circle, yellow square, and purple triangle represent the measured angles for topological charges of 60, 80, and 100, respectively, while the lines represent the linear fit of the measured angles.}
\end{figure*}

The angle variation is measured, taking into account the LS hotspot. This means that the angle extracted is the clockwise angle on the x-axis (Fig. \ref{fig:AngleExt}a) of the LS, considering its weighted intensity hotspot (Fig. \ref{fig:AngleExt}b). The radius extraction is done considering the central $\ell$ applied to create the LS. This value is used to calculate both the experimental $v_{og}$ (Eq. \ref{EqOGV1}) and the programmed $v_{og}$ (Eq. \ref{EqOGV2}).

Since the $v_{og}$ tunability is explored in this work via the manipulation of the LS radius, the $\frac{\partial \omega}{\partial l}$, and therefore the $\frac{\partial \alpha(t)}{\partial t}$ should be the same to all topological charges used to generate the LS. This can be seen in Fig. \ref{fig:AngleExt}b, the range of topological charges used to synthesize the LS is different, but the experimentally measured slope is approximately the same.

\section{Light spring simulations}

To simulate the LS propagation, a simulation box was defined by a $2D$ window with a given $X-Y$ dimension, discretized into $M=100$ pixels, with a time interval corresponding to the LS propagation through that window. The time interval was divided into an integer number of $N$ steps (typically around 1200), each time step determined by an $M \times M$ matrix of the LS cross-section at that corresponding time. This results in an $M \times M \times N$ complex matrix of the calculated electric field of the LS. 

The $X-Y$ dimension of the window and time interval were adjusted according to the beam waist and the $v_{og}$ value to ensure an accurate and efficient simulation of the experimental conditions. Each simulated LS was defined by the frequency bandwidth (defined by a low-end and high-end value), the applied topological charges (defined by a low-end integer value and a high-end integer value), and the number of spectral windows (an integer value determined by the spatiospectral linear coupling that generates the LS). 

The LS was generated from the superposition of simulated Bessel-Gauss (BG) beams with a fixed beam waist. For each time step, the cross-sectional electric field was calculated by the sum of the BG electric fields corresponding to the input parameters used to create the LS, according to Eq. 4 from \cite{SimulationLS_SI}. After the LS was simulated, the $v_{og}$ and the LS radius were extracted from it and compared with the measured experimental values. 

\section{Superradiant plasma mechanism simulations description}
\label{sec:PIC}

The simulations of the interaction between the light springs and the overdense plasma were performed using the PIC code Osiris \cite{OSIRIS_SI} with the radiation diagnostic tool Radio \cite{RaDiO_SI} turned on. 

To simulate the LS in the PIC code, we decompose the field into Laguerre-Gaussian (LG) modes rather than Bessel-Gauss (BG) modes. This decision is motivated by the finite transverse size of LG modes; compared to similar BG modes, LG modes allow for significantly smaller simulation domains. This choice does not impact the final conclusions, as the underlying physical mechanism relies solely on the orbital group velocity. This velocity depends only on the radial position of the intensity maximum and the rotation speed, which is determined by the frequency-dependent azimuthal mode indices.

The LS propagates in the positive $z$ direction and is linearly polarized in the $y$ direction. It is composed of 31 Laguerre-Gaussian (LG) modes with frequencies evenly spaced in the range $\omega\in[ 0.553; 0.703 ]\omega_p$ (average frequency $\omega_0=0.628\omega_p$ with spacing $\Delta\omega=0.005\omega_p$). The radial index of the LG modes is set to $p=0$ and the azimuthal index is set to achieve the desired orbital group velocity: consecutive LG modes differ by $\Delta\ell=1$ and the average index is $\ell_0$. The parameters used can be seen in Table 1 in the main text.
Each LG mode has a spot size of $50 c/ \omega_p$, focuses at the origin, has a longitudinal squared sine envelope with rise and fall lengths of $500 c/ \omega_p$ and a flat length of $4000 c/ \omega_p$, and equal normalized peak wave vector amplitude. These parameters allow both LS to have the same energy.

The intensity of each Laguerre-Gaussian mode is given by
\begin{equation}
    |E_y(z,r,t)|=E_0\frac{w_0}{w(z)}\left(\frac{r\sqrt2}{w(z)}\right)^{|\ell|}\exp\left(\frac{-r^2}{w^2(z)}\right),
\end{equation}
where $ z_R=\pi w_0^2/\lambda_0 $ and $ w(z)=w_0\sqrt{1+(z/z_R)^2} $. The beam divergence angle is then given by
\begin{equation}
    \theta_{BD} = \lim_{z\to\infty}\arctan\frac{r_{\max}(z)}{z} = \arctan\frac{\sqrt{\ell_0}\lambda_0}{\sqrt{2}\pi w_0}
\end{equation}
where $r_{\max}(z)$ satisfies $ ( \partial_r |E_y| )_{r=r_{\max}}=0 $.

The electron plasma slab ($m/q=-1$) has a density of $ n_0=1 $ and is confined in the $z$ direction between $z=\pm10.0c/\omega_p$. 

The radiation from $20\%$ of the particles is captured in a spherical detector positioned in the far field in the negative $z$ direction, at $z=-10^{6}c/\omega_p$. It captures the radiation up to $0.8\text{ rad}$ in all directions with respect to the negative $z$ axis between $t=-500\omega_p^{-1}$ and $t=5500\omega_p^{-1}$. It is composed of $140$ cells in each angular direction and $4000$ cells in the temporal domain.

The PIC simulation box ranges from $z=-30.0c/\omega_p$ to $z=30.0c/\omega_p$, with $n_z=115$ cells, and from $x,y=-800.0c/\omega_p$ to $x,y=800.0c/\omega_p$, with $n_{x,y}=600$ cells. Each cell contains 4 simulation particles. The simulation is run from $t=0.0\omega_p^{-1}$ to $t=5500.0\omega_p^{-1}$ in steps $dt=0.5\omega_p^{-1}$. Both the electromagnetic fields and the particles have open boundary conditions in all spatial dimensions.
% dt_courant = 0,05178382224

Using a normalized plasma density of $n_0=2.824\times10^{21}\ cm^{-3}$, both LS have a wavelength of $\lambda_0=1\ \mu m$, a spot size of $ w_0=5\mu m $, a flat temporal duration of $\tau\approx 1667.81fs$ and a total energy of $ 0.5\ J $.

\section{Superradiant plasma mechanism: simulation results}

\subsection{Temporal description of the train of unipolar pulses}

\begin{figure*}[ht]
\includegraphics[scale=1]{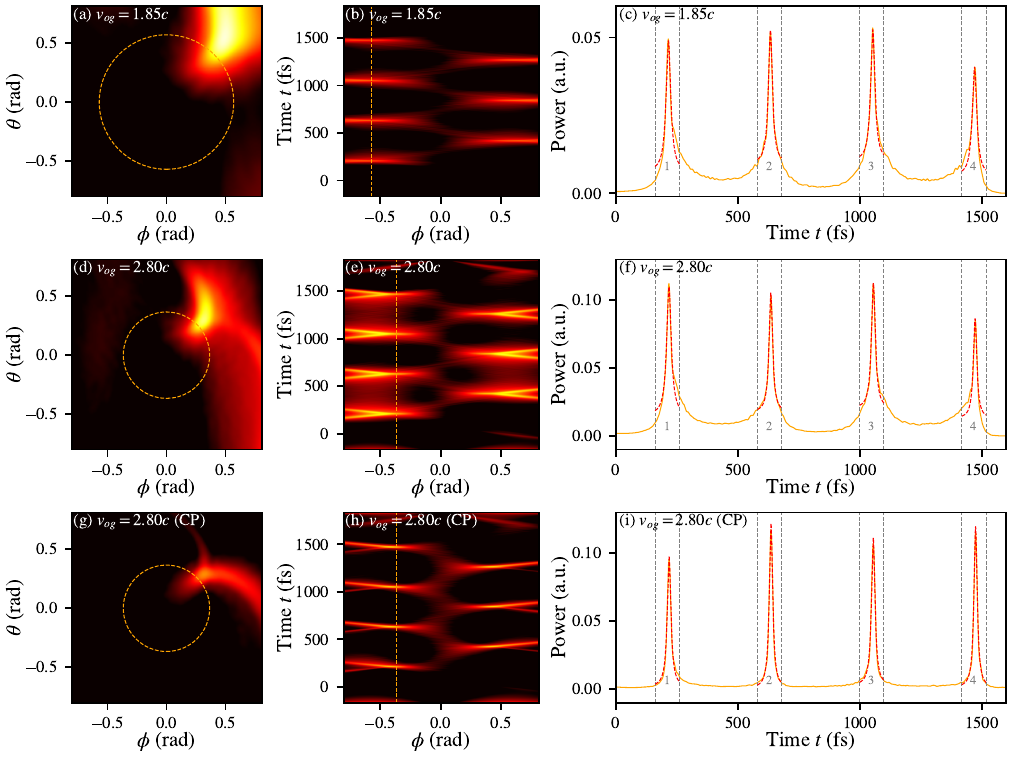}
\caption{\label{fig:temporal_fit} \textbf{Temporal analysis and Lorentzian fitting of superradiant THz LS emission} (a, d, g) Instantaneous radiation patterns on the detector at different group velocities $v_{og}$ and polarizations (LP: Linear, CP: Circular). The dashed yellow circles indicate the region of superradiant emission. (b, e, h) Temporal evolution of the radiation at $\theta=0$ as a function of the azimuthal angle $\phi$. (c, f, i) Temporal profiles of the four distinct radiation bursts extracted at $\phi=-\theta_{CH}$. The orange solid lines represent the simulated data, while the red dashed lines show the Lorentzian fits. }
\end{figure*}

The temporal evolution of the light bursts in the superluminal case are here further analyzed. Fig.~\ref{fig:temporal_fit}~a, d and g show the instantaneous radiation on the detector filtered only for the frequencies of the superradiant THz LS. Fig.~\ref{fig:temporal_fit}b, e and h show the temporal evolution of the same filtered radiation on the detector at $\theta=0$. The four radiation bursts at $\phi=-\theta_{CH}$ are then plotted on Fig.~\ref{fig:temporal_fit}c, f and i. These peaks were fitted to Lorentzian functions given by
\[
f(t) = A_0\frac{(\Gamma/2)^2}{(t-t_0)^2+(\Gamma/2)^2}.
\]
The results are shown in Table~\ref{tab:SimLSFits}. All have a similar temporal duration of $\approx20fs$ and are spaced by $\approx419fs$, exactly the temporal pitch of the light springs.

\begin{table}[h!]
\centering
\setlength{\tabcolsep}{5pt}
\begin{tabular}{ ccccccc }
    \hline\hline
    $v_{og}$ (c) & Pol. & Peak & $A_0$ (a.u.) & $\Gamma$ (fs) & $t_0$ (fs) & $\Delta t$ (fs) \\ 
    \hline
    1.85 & LP & 1 & 0.042 & 22.852 &  214.820 &        - \\
    1.85 & LP & 2 & 0.043 & 19.502 &  632.941 &  418.120 \\
    1.85 & LP & 3 & 0.042 & 20.950 & 1052.711 &  419.770 \\
    1.85 & LP & 4 & 0.035 & 22.167 & 1469.982 &  417.271 \\
    \hline
    2.80 & LP & 1 & 0.095 & 22.453 &  217.036 &        - \\
    2.80 & LP & 2 & 0.088 & 17.641 &  634.646 &  417.610 \\
    2.80 & LP & 3 & 0.093 & 18.940 & 1055.087 &  420.440 \\
    2.80 & LP & 4 & 0.074 & 17.127 & 1472.224 &  417.137 \\
    \hline
    2.80 & CP & 1 & 0.097 & 17.666 &  217.466 &        - \\
    2.80 & CP & 2 & 0.122 & 16.419 &  635.370 &  417.904 \\
    2.80 & CP & 3 & 0.109 & 17.105 & 1054.596 &  419.226 \\
    2.80 & CP & 4 & 0.119 & 15.983 & 1474.261 &  419.665 \\
    \hline\hline
\end{tabular}
\caption{\label{tab:SimLSFits} Lorentzian fitting parameters for the peaks in Fig~\ref{fig:temporal_fit} of main text. }
\end{table}

\subsection{Wavefront picture}

To better understand the quasiparticle picture, let us consider that the radiation is generated by azimuthally evenly spaced emitters at the LS radius, $r_{LS}$, radiating unipolar spherical waves with a delay given by $\Delta t=r_{LS}\Delta\alpha/v_{og}$, where $\Delta\alpha$ is the azimuthal spacing of the emitters. The wavefronts from each wave will appear as a line in the detector, as seen in Figures~4(b) and (c) of the main text. For the superluminal case, it is clear that these lines superimpose each other at the Cherenkov angle, creating the characteristic cardioid shape. The discrepancies between the theoretical lines and the simulated intensity arise from the fact that the radiation is not composed by a single peaked wavefront but a more smoothed out wave, which blurs the intensity on the detector.

To derive the lines on the detector, consider a spherical detector with radius $R_{d}$ centered at the origin is described in spherical coordinates $(\theta,\phi)$ as 
\[
\begin{cases} 
z_d = R_{d}\sin\theta\cos\phi, \\ 
x_d = R_{d}\sin\theta\sin\phi, \\ 
y_d = R_{d}\cos\theta 
\end{cases}
\] 
A spherical wave composed by a single wavefront is radiated by an emitter located at $ \mathbf{r}=(z_0,x_0,y_0) $ at $ t=0 $. If this wavefront hits the detector at a time $ t $, then the intersection is given by:
\[
\theta(\phi)=\arctan\!\left(\frac{y_0^{2}A \pm B \sqrt{y_0^{2}C}}{y_0\big(AB \mp \sqrt{y_0^{2}C}\big)}\right),
\]
where
\[
\begin{aligned}
A &= R_{d}^{2} + |\mathbf{r}|^{2} -c^2t^{2},\\
B &= z_{0}\cos\phi + x_{0}\sin\phi,\\
C &= 2R_{d}^{2}y_{0}^{2} + 2c^{2}t^{2}\big(R_{d}^{2} + |\mathbf{r}|^{2}\big)
    -c^{4}t^{4} - R_{d}^{4} - |\mathbf{r}|^{4} - 2R_{d}^{2}\Big((x_{0}^2-z_{0}^2)\cos2\phi
    - 2x_{0}z_{0}\sin2\phi\Big).
\end{aligned}
\]
This equation was used to create Fig.4b and c in the main text and the movie by considering 600 emitters located at $\mathbf{r_i}=(z_0,r_{LS}\cos\alpha_i,r_{LS}\sin\alpha_i) $ emitting at $t_i=r_{LS}\alpha_i/v_{og}$, with $\alpha_i$ uniformly distributed in the range $[0,6\pi]$.

\subsection{Quasiparticle shape factor and radiation asymetry}

The radiation of a quasiparticle \cite{SuperradianceQuasiparticle_SI} per unit solid angle per unit frequency in the far-field is given by
\begin{equation}
    \frac{d^2 I}{d \omega d \Omega}=\frac{\omega^2}{4 \pi^2 c^3}|\boldsymbol{\mathcal { S }}(\omega, \Omega)|^2\left|\int d \tau \exp \left[i \omega\left(\tau-\mathbf{n} \cdot \mathbf{r}_{\mathbf{c}}(\tau) / c\right)\right]\right|^2,
\end{equation}
where $\tau$ is the time of emission, $r_c(\tau)$ is the position of the quasiparticle, $\omega$ is the radiation frequency, $\Omega$ is the solid angle, $\mathbf{n}$ is a unit vector that sets the observation direction and $\boldsymbol{\mathcal { S }}(\omega, \Omega)$ is the quasiparticle shape factor, defined as 
\begin{equation}
    \boldsymbol{\mathcal{S}(\omega, \Omega)}=\mathbf{n} \times\left[\mathbf{n} \times \tilde{\mathbf{j}}\left(\mathbf{k}_{\xi}\right)\right],
\end{equation}
with $\tilde{\mathbf{j}}\left(\mathbf{k}_{\xi}\right)$ the Fourier transform of the current of the quasiparticle and $\mathbf{k}_{\xi}\equiv\omega\mathbf{n}/c$.

\begin{figure*}[ht]
\includegraphics[scale=1]{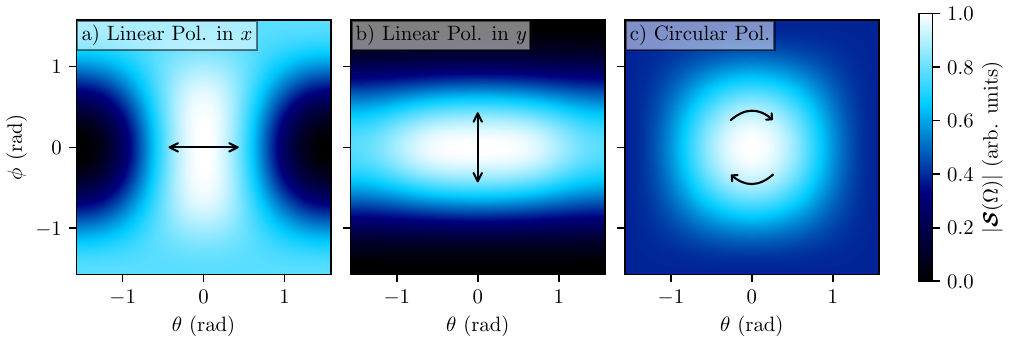}
\caption{\label{fig:ls_asymetry} \textbf{Quasiparticle shape factor} plotted in the detector angles for 3 different polarization states. }
\end{figure*}

In this work, we can approximate the quasiparticle shape as a gaussian with a standard deviation of $\sigma\approx10\mu m$ (estimative) and thus
\begin{equation*}
    \tilde{\mathbf{j}}\left(\mathbf{k}_{\xi}\right) = \tilde{j_0} \exp\left(\frac{-|\mathbf{k}_{\xi}|^2\sigma^2}{2}\right) \mathbf{\hat e_{\rm pol}}
\end{equation*}
with $\tilde{j_0}$ an amplitude and $\mathbf{\hat e_{\rm pol}}$ a unit vector that defines polarization ($\mathbf{\hat e_{\rm pol}}=\mathbf{\hat x}$ for linear in $x$, $\mathbf{\hat e_{\rm pol}}=\mathbf{\hat y}$ for linear in $y$ and $\mathbf{\hat e_{\rm pol}}=\mathbf{\hat x}+i\mathbf{\hat y}$ for circular). Using the coordinate system defined as $\mathbf{n}=(n_x,n_y,n_z)=(\cos\phi\sin\theta, \sin\phi, \cos\phi\cos\theta)$, the shaped factor can be calculated. Figure~\ref{fig:ls_asymetry} portrays this factor for all the relevant polarizations, yielding asymmetric profiles for linear polarizations. This explains the asymmetric radiation profile when using a linearly polarized LS as the plasma currents will have the same polarization as the driver laser pulse.

\subsection{Spectral phase}

\begin{figure*}[h!]
\includegraphics[scale=0.7]{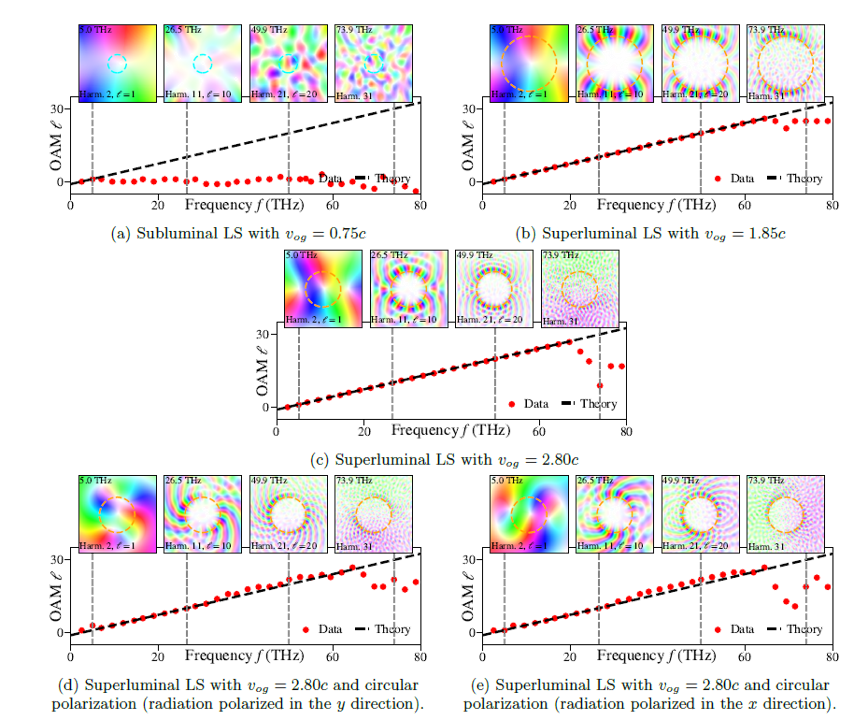}
\caption{\label{fig:spectralphase} \textbf{Spectral phase and Orbital Angular Momentum (OAM) analysis}. The top plots show the spectral phase on the detector for four different frequencies in the THz range. The bottom plots portray the extracted OAM topological order at the Cherenkov angle (beam LS pitch. At each harmonic, the phase wraps around an integer number of times around the axis. }
\end{figure*}

To assess the orbital angular momentum content of the radiation coming from the LS interaction with the plasma, the spectral intensity and phase were retrieved from the electric field parallel to the LS polarization direction pattern. Figure~\ref{fig:spectralphase} shows the result of this study. For each $n^{th}$ harmonic, the phase wraps around $n-1$ times angularly, corresponding to a azimuthal index $n$. The shift $-1$ comes from the polarization of the radiation. The correspondence between the theoretical value and the retrieved OAM index is very good for high harmonic numbers in the superluminal case. On the other hand, for the subluminal case, the spectral phase is random after a few harmonics and thus the radiation pattern is not a clear THz LS.

\subsection{Efficiency estimation}

To estimate the efficiency of the conversion process, the magnetic field at a perpendicular slice to the plasma near the surface of the plasma was captured in two simulations: without and with the plasma. The first one captured the incident radiation while the difference between the second and the first captured the reflected radiation. The reflected radiation was separated in 3 different ranges: the first between 264 and 355 THz corresponding to the driving frequency range; the second between 2 and 72 THz that encompasses the superradiant THz LS; and the third between 528 and 671 THz that corresponds to the second harmonic of the radiation. We chose to study the magnetic and not the electric field as a way to better isolate the purely electromagnetic components while filtering out the electrostatic that may arise from the plasma surface.

\begin{figure*}[h!]
\includegraphics[scale=0.9]{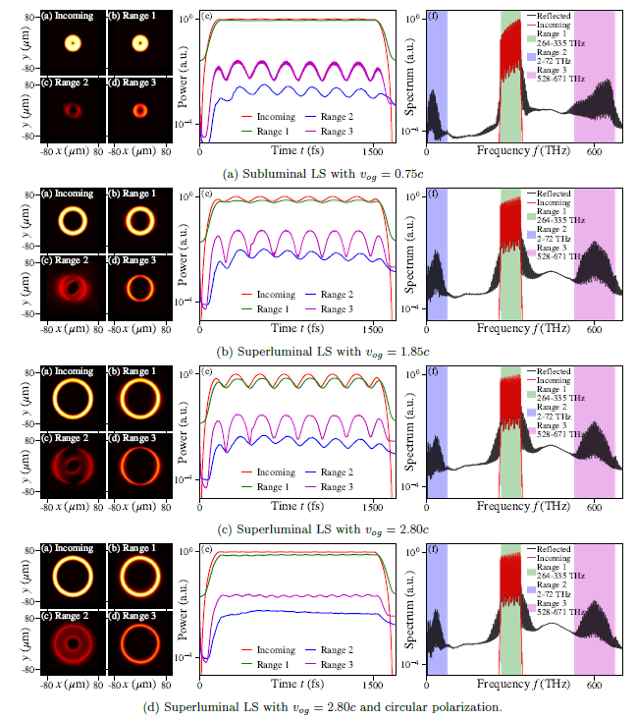}
\caption{\label{fig:efficiency} Efficiency of the 3 main components of the reflected radiation captured on the grid of the simulation close to the plasma surface.}
\end{figure*}

The details of the analysis are plotted in Figures~\ref{fig:efficiency}~a-d. The plots on the left of each subfigure represent the total integrated energy density over the detector plane for the incoming radiation and the 3 frequency ranges separately. The middle plots show the total integrated energy density correspondent to each of the components measured. The right plots correspond to the incoming driver spectrum (red line) and the total reflected radiation captured by the detector (black line).

By integrating the power on the drive frequency range in the incident spectrum and in the 3 different frequency ranges in the reflected spectrum, the efficiencies for each of the reflected components can be obtained. The results are portrayed in Table~\ref{tab:efficiency}.

\begin{table}[h!]
\centering
\setlength{\tabcolsep}{5pt}
\begin{tabular}{ ccccc }
    \hline\hline
    $v_{og}$ ($c$) & Pol. & Range 1 (\%) & Range 2 (\%) & Range 3 (\%) \\ 
    \hline
    % electric field
    %0.75 & LP & 91.102 & 0.066 & 1.454 \\
    %1.85 & LP & 83.559 & 0.225 & 3.528 \\
    %2.80 & LP & 82.684 & 0.254 & 2.781 \\
    %2.80 & CP & 81.725 & 0.261 & 2.748 \\
    % magnetic field
    0.75 & LP & 91.034 & 0.189 & 1.345 \\
    1.85 & LP & 81.367 & 0.727 & 3.443 \\
    2.80 & LP & 76.677 & 0.437 & 2.449 \\
    2.80 & CP & 78.117 & 0.518 & 2.213 \\
    \hline\hline
\end{tabular}
\caption{\label{tab:efficiency} Efficiency of the 3 main components of the reflected radiation captured on the grid of the simulation close to the plasma surface. }
\end{table}

%%% Citations %%%
%


\begin{thebibliography}{}

\end{thebibliography}


\begin{thebibliography}{51}%

\bibitem{STNoise}
S.~W.~Jolly, O.~Gobert, and F.~Quéré,
``Spatio-temporal characterization of ultrashort laser beams: a tutorial,''
\emph{Journal of Optics} \textbf{22}, 103501 (2020).

\bibitem{STAberration}
D.~Brinks, R.~Hildner, F.~D.~Stefani, and N.~F.~van Hulst,
``Beating spatio-temporal coupling: implications for pulse shaping and coherent control experiments,''
\emph{Optics Express} \textbf{19}, 26486 (2011).

\bibitem{STBeams}
Y.~Shen \emph{et al.},
``Roadmap on spatiotemporal light fields,''
\emph{Journal of Optics} \textbf{25}, 093001 (2023).

\bibitem{STBeams1}
D.~Cruz-Delgado \emph{et al.},
``Synthesis of ultrafast wavepackets with tailored spatiotemporal properties,''
\emph{Nature Photonics} (2022).

\bibitem{STMultimodeShaping}
M.~Piccardo \emph{et al.},
``Roadmap on multimode light shaping,''
\emph{Journal of Optics} \textbf{24}, 013001 (2021).

\bibitem{NLConversion1}
A.~de~las~Heras \emph{et al.},
``Extreme-ultraviolet vector-vortex beams from high harmonic generation,''
\emph{Optica} \textbf{9}, 71 (2021).

\bibitem{NLConversion2}
L.~Rego \emph{et al.},
``Generation of extreme-ultraviolet beams with time-varying orbital angular momentum,''
\emph{Science} \textbf{364} (2019).

\bibitem{QuantumMetrology}
F.~S.~Roux,
``Nonlinear interferometry in all spatiotemporal degrees of freedom,''
\emph{Phys. Rev. A} \textbf{105} (2022).

\bibitem{InvariantSTBeams}
M.~Yessenov, B.~Bhaduri, H.~E.~Kondakci, and A.~F.~Abouraddy,
``Classification of propagation-invariant space-time wave packets in free space: Theory and experiments,''
\emph{Phys. Rev. A} \textbf{99} (2019).

\bibitem{almeida_universal_2025}
R.~Almeida, D.~Ramsey, A.~F.~Abouraddy, J.~P.~Palastro, and J.~Vieira,
``Universal structure of propagation-invariant optical pulses,''
\emph{Opt. Lett.} \textbf{50} (2025).

\bibitem{STOV}
A.~Bekshaev,
``Spatiotemporal optical vortices: Principles of description and basic properties,''
\emph{APL Photonics} \textbf{9} (2024).

\bibitem{AbouraddyGV}
H.~E.~Kondakci and A.~F.~Abouraddy,
``Optical space-time wave packets having arbitrary group velocities in free space,''
\emph{Nature Communications} \textbf{10} (2019).

\bibitem{STwavepackets}
M.~Yessenov \emph{et al.},
``Space-time wave packets,''
\emph{Advances in Optics and Photonics} \textbf{14}, 455 (2022).

\bibitem{sainte-marie_controlling_2017}
A.~Sainte-Marie, O.~Gobert, and F.~Quéré,
``Controlling the velocity of ultrashort light pulses in vacuum through spatio-temporal couplings,''
\emph{Optica} \textbf{4} (2017).

\bibitem{Froula_PoP_2017}
D.~H.~Froula \emph{et al.},
``Flying focus: Spatial and temporal control of intensity for laser-based applications,''
\emph{Physics of Plasmas} \textbf{26}, 032109 (2019).

\bibitem{FlyingFocus}
A.~M.~V. \emph{et al.},
``Programmable-trajectory ultrafast flying focus pulses,''
\emph{Optics Express} \textbf{31}, 31354 (2023).

\bibitem{kabacinski_spatio-temporal_2023}
A.~Kabacinski, E.~Oliva, F.~Tissandier, J.~Gautier, M.~Kozlová, J.-P.~Goddet,
I.~A.~Andriyash, C.~Thaury, P.~Zeitoun, and S.~Sebban,
``Spatio-temporal couplings for controlling group velocity in longitudinally pumped seeded soft X-ray lasers,''
\emph{Nat. Photon.} \textbf{17} (2023).

\bibitem{liberman_use_2024}
A.~Liberman, R.~Lahaye, S.~Smartsev, S.~Tata, S.~Benracassa, A.~Golovanov,
E.~Levine, C.~Thaury, and V.~Malka,
``Use of spatiotemporal couplings and an axiparabola to control the velocity of peak intensity,''
\emph{Opt. Lett.} \textbf{49} (2024).

\bibitem{markland_rapidly_2026}
H.~S.~Markland, J.~Rosenbluth, R.~Boni, M.~V.~Ambat, C.~Dorrer,
J.~P.~Palastro, D.~H.~Froula, and J.~J.~Pigeon,
``Rapidly tunable ultrabroadband flying focus using adaptive optics and an axiparabola,''
\emph{Opt. Lett.} \textbf{51} (2026).

\bibitem{MS}
A.~M.~Shaltout, V.~M.~Shalaev, and M.~L.~Brongersma,
``Spatiotemporal light control with active metasurfaces,''
\emph{Science} \textbf{364} (2019).

\bibitem{MS2}
L.~Chen \emph{et al.},
``Synthesizing ultrafast optical pulses with arbitrary spatiotemporal control,''
\emph{Science Advances} \textbf{8} (2022).

\bibitem{Rosales_SLM}
C.~Rosales-Guzmán and A.~Forbes,
``How to shape light with spatial light modulators,''
in \emph{How to Shape Light with Spatial Light Modulators}
(Springer, 2017), Chap.~4, pp.~11--30.

\bibitem{SynthesisSheets}
M.~A.~Romer, L.~A.~Hall, and A.~F.~Abouraddy,
``Synthesis and characterization of space-time light sheets: a tutorial,''
\emph{Journal of Optics} (2024).

\bibitem{QuereLS}
G.~Pariente and F.~Quéré,
``Spatio-temporal light springs: extended encoding of orbital angular momentum in ultrashort pulses,''
\emph{Optics Letters} \textbf{40}, 2037 (2015).

\bibitem{MarcoLS}
M.~Piccardo \emph{et al.},
``Broadband control of topological--spectral correlations in space--time beams,''
\emph{Nature Photonics} \textbf{17}, 822--828 (2023).

\bibitem{LS2}
M.~de~Oliveira and A.~Ambrosio,
``Subcycle modulation of light’s orbital angular momentum via a Fourier space-time transformation,''
\emph{Science Advances} \textbf{11} (2025).

\bibitem{LS3}
Q.~Lin \emph{et al.},
``Direct space--time manipulation mechanism for spatio-temporal coupling of ultrafast light field,''
\emph{Nature Communications} \textbf{15} (2024).

\bibitem{SyntMotion1}
A.~C.~Harwood \emph{et al.},
``Space-time optical diffraction from synthetic motion,''
\emph{Nature Communications} \textbf{16} (2025).

\bibitem{SyntMotion2}
F.~Belgiorno \emph{et al.},
``Quantum radiation from superluminal refractive-index perturbations,''
\emph{Phys. Rev. Lett.} \textbf{104}, 140403 (2010).

\bibitem{SyntMotion3}
J.~Sloan \emph{et al.},
``Controlling two-photon emission from superluminal and accelerating index perturbations,''
\emph{Nature Physics} \textbf{18}, 67--74 (2021).

\bibitem{Attosecond}
P.~Agostini and L.~F.~DiMauro,
``The physics of attosecond light pulses,''
\emph{Rep. Prog. Phys.} \textbf{67}, 813--855 (2004).

\bibitem{RadiusOAM}
A.~M.~Yao and M.~J.~Padgett,
``Orbital angular momentum: origins, behavior and applications,''
\emph{Advances in Optics and Photonics} \textbf{3}, 161 (2011).

\bibitem{SolidState}
T.~Omatsu and S.~R.~Allam,
``Structured light solid-state laser sources,''
\emph{Journal of Optics} \textbf{27}, 073001 (2025).

\bibitem{JorgePlasmaAcceleration}
J.~Vieira, J.~T.~Mendonça, and F.~Quéré,
``Optical control of the topology of laser-plasma accelerators,''
\emph{Phys. Rev. Lett.} \textbf{121} (2018).

\bibitem{VieiraHighOrbitalAngular2016}
J.~Vieira \emph{et al.},
``High orbital angular momentum harmonic generation,''
\emph{Phys. Rev. Lett.} \textbf{117}, 265001 (2016).

\bibitem{OSIRIS}
R.~A.~Fonseca \emph{et al.},
``OSIRIS: A three-dimensional, fully relativistic particle-in-cell code for modeling plasma-based accelerators,''
in \emph{Computational Science --- ICCS 2002},
edited by P.~M.~A.~Sloot \emph{et al.}
(Springer, Berlin, Heidelberg, 2002), pp.~342--351.

\bibitem{RaDiO}
M.~Pardal \emph{et al.},
``RaDiO: An efficient spatiotemporal radiation diagnostic for particle-in-cell codes,''
\emph{Computer Physics Communications} \textbf{285}, 108634 (2023).

\bibitem{LichtersShortpulseLaserHarmonics1996}
R.~Lichters, J.~Meyer-ter Vehn, and A.~Pukhov,
``Short-pulse laser harmonics from oscillating plasma surfaces driven at relativistic intensity,''
\emph{Physics of Plasmas} \textbf{3}, 3425--3437 (1996).

\bibitem{DenoeudInteractionUltraintenseLaser2017}
A.~Denoeud \emph{et al.},
``Interaction of ultraintense laser vortices with plasma mirrors,''
\emph{Phys. Rev. Lett.} \textbf{118}, 033902 (2017).

\bibitem{LeblancPlasmaHologramsUltrahighintensity2017}
A.~Leblanc \emph{et al.},
``Plasma holograms for ultrahigh-intensity optics,''
\emph{Nature Physics} \textbf{13}, 440--443 (2017).

\bibitem{Superradiance2}
J.~Vieira \emph{et al.},
``Generalized superradiance for producing broadband coherent radiation with transversely modulated arbitrarily diluted bunches,''
\emph{Nature Physics} \textbf{17}, 99--104 (2020).
 
\bibitem{SuperradianceQuasiparticle}
B.~Malaca \emph{et al.},
``Coherence and superradiance from a plasma-based quasiparticle accelerator,''
\emph{Nature Photonics} \textbf{18}, 39--45 (2023).

\bibitem{THzFlyingFocus}
S.~Fu, B.~Groussin, Y.~Liu, A.~Mysyrowicz, V.~Tikhonchuk, and A.~Houard,
``Steering Laser-Produced THz Radiation in Air with Superluminal Ionization Fronts,''
\emph{Phys. Rev. Lett.} \textbf{134} (2025).

\bibitem{THzTwoColor}
T.~T.~Simpson, J.~J.~Pigeon, M.~V.~Ambat, K.~G.~Miller, D.~Ramsey,
K.~Weichman, D.~H.~Froula, and J.~P.~Palastro,
``Spatiotemporal control of two-color terahertz generation,''
\emph{Phys. Rev. Res.} \textbf{6} (2024).

\bibitem{THzCherenkov}
L.~A.~Johnson, J.~P.~Palastro, T.~M.~Antonsen, and K.~Y.~Kim,
``THz generation by optical Cherenkov emission from ionizing two-color laser pulses,''
\emph{Phys. Rev. A} \textbf{88} (2013).

\bibitem{pardal_SuperradiantScatteringEvanescent2026}
M.~Pardal, R.~A.~Fonseca, and J.~Vieira,
``Superradiant Scattering from Evanescent Waves,''
\emph{Phys. Rev. Res.} \textbf{8} (2026).

\bibitem{jang_EfficientTerahertzBrunel2019}
D.~Jang \emph{et al.},
``Efficient terahertz and Brunel harmonic generation from air plasma via mid-infrared coherent control,''
\emph{Optica} \textbf{6}, 1338 (2019).

\bibitem{tailliez_TerahertzPulseGeneration2020}
C.~Tailliez \emph{et al.},
``Terahertz pulse generation by two-color laser fields with circular polarization,''
\emph{New Journal of Physics} \textbf{22}, 103038 (2020).

\bibitem{pak2023multi}
T.~Pak \emph{et al.},
``Multi-millijoule terahertz emission from laser-wakefield-accelerated electrons,''
\emph{Light: Science \& Applications} \textbf{12}, 37 (2023).

\bibitem{liao2019multimillijoule}
G.~Liao \emph{et al.},
``Multimillijoule coherent terahertz bursts from picosecond laser-irradiated metal foils,''
\emph{Proc. Natl. Acad. Sci. USA} \textbf{116}, 3994--3999 (2019).

\bibitem{HighPower}
B.~Oliveira \emph{et al.},
``High-aspect-ratio, ultratall silica meta-optics for high-intensity structured light,''
\emph{Optica} \textbf{12}, 713 (2025).

\end{thebibliography}

\begin{thebibliography}{5}%

\bibitem{AbouraddyGV_SI}
H.~E.~Kondakci and A.~F.~Abouraddy,
``Optical space-time wave packets having arbitrary group velocities in free space,''
\emph{Nature Communications} \textbf{10} (2019).

\bibitem{SimulationLS_SI}
H.~Zhang, J.~Zeng, X.~Lu, Z.~Wang, C.~Zhao, and Y.~Cai,
``Review on fractional vortex beam,''
\emph{Nanophotonics} \textbf{11}, 241--273 (2021).

\bibitem{OSIRIS_SI}
R.~A.~Fonseca \emph{et al.},
``OSIRIS: A three-dimensional, fully relativistic particle-in-cell code for modeling plasma-based accelerators,''
in \emph{Computational Science --- ICCS 2002},
edited by P.~M.~A.~Sloot \emph{et al.}
(Springer, Berlin, Heidelberg, 2002), pp.~342--351.

\bibitem{RaDiO_SI}
M.~Pardal \emph{et al.},
``RaDiO: An efficient spatiotemporal radiation diagnostic for particle-in-cell codes,''
\emph{Computer Physics Communications} \textbf{285}, 108634 (2023).

\bibitem{SuperradianceQuasiparticle_SI}
B.~Malaca \emph{et al.},
``Coherence and superradiance from a plasma-based quasiparticle accelerator,''
\emph{Nature Photonics} \textbf{18}, 39--45 (2023).

\end{thebibliography}
\end{document}